\documentclass[aps,nofootinbib,preprint,superscriptaddress]{revtex4}%
\usepackage{hyperref}
\usepackage{amsmath}
\usepackage{amsfonts}
\usepackage{amssymb}
\usepackage{graphicx}
\usepackage{color}%
\providecommand{\U}[1]{\protect\rule{.1in}{.1in}}

\begin{document}
\title{Comment on "High-Energy Symmetry of String Theory" }
\author{Sheng-Hong Lai}
\email{xgcj944137@gmail.com}
\affiliation{Department of Electrophysics, National Yang Ming Chiao Tung University,
Hsinchu, Taiwan, R.O.C.}
\affiliation{Department of Electrophysics, National Chiao-Tung University, Hsinchu, Taiwan, R.O.C.}
\author{Jen-Chi Lee}
\email{jcclee@cc.nctu.edu.tw}
\affiliation{Department of Electrophysics, National Yang Ming Chiao Tung University,
Hsinchu, Taiwan, R.O.C.}
\affiliation{Department of Electrophysics, National Chiao-Tung University, Hsinchu, Taiwan, R.O.C.}
\author{Yi Yang}
\email{yiyang@mail.nctu.edu.tw}
\affiliation{Department of Electrophysics, National Yang Ming Chiao Tung University,
Hsinchu, Taiwan, R.O.C.}
\affiliation{Department of Electrophysics, National Chiao-Tung University, Hsinchu, Taiwan, R.O.C.}
\date{\today}

\begin{abstract}
Following the recent discovery of the Lauricella string scattering amplitudes
(LSSA) and their associated exact $SL(K+3,C)$ symmetry, we give a brief
comment on Gross conjecture regarding "High-energy symmetry of string theory".

\end{abstract}
\maketitle

The \textit{Gross conjecture} regarding high-energy symmetry of string theory
\cite{Gross,Gross1} was based on the saddle-point calculation of hard string
scattering amplitudes (SSA) of both the closed \cite{GM,GM1} and open
\cite{GrossManes} string theories. The conjecture claimed that there existed
infinite \textit{linear relations} with constant\textit{ }coefficients\textit{
}among hard SSA of different string states. Moreover, these infinite linear
relations were so powerful that they can be used to solve all the hard SSA and
express them in terms of one amplitude. Some monographs \cite{Polchin,Kaku}
had made speculations about this hidden stringy symmetry without getting any
conclusive results. However, the saddle-point calculation of the hard SSA
\cite{GM1,GrossManes} which was claimed to be valid for all string states and
all string loop orders was pointed out to be inconsistent for the cases of the
\textit{excited} string states in a series of works in
\cite{ChanLee,ChanLee2,CHL}.

It was then further shown \cite{Closed} that even at closed string-tree level,
there was \textit{no} reliable saddle-point in the hard SSA calculation. The
authors of \cite{Closed} gave three evidences to demonstrate the inconsistency
of the \textit{saddle-point}. So instead of using the saddle-point method,
they first used the infinite linear relations (to be discussed in
Eq.(\ref{04}) below) to derive the \textit{hard string BCJ relations
}\cite{bcj0} and then used the KLT \cite{KLT} formula to obtain the correct
hard closed SSA, which differs from result of Gross and Mende \cite{GM1} by an
oscillation prefactor. This prefactor consistently implied the existence of
infinitely many zeros and poles in the hard SSA.

Soon later a similar conclusion was independently made in \cite{West2}, which
was based on the group theoretical calculation \cite{NW5} of SSA. The Authors
of \cite{West2} found out that up to the string one-loop level the
saddle-point calculation was valid only for the hard four tachyon SSA, but was
incorrect for other hard SSA of excited string states. For this reason, the
authors of \cite{West2} admitted that they can not consistently find out any
linear relations as suggested in \cite{Gross}.

For the case of open bosonic string at the mass level $M^{2}=4$, as an
example, the linear relations (without calculating the ratios) of hard open
SSA of Gross and Manes \cite{GrossManes} were miscalculated to be
\cite{GrossManes}%
\begin{equation}
\mathcal{T}_{TTT}\propto\mathcal{T}_{[LT]}\text{ \ \ , \ }\mathcal{T}%
_{LLT}=\mathcal{T}_{(LT)}=0,\label{d}%
\end{equation}
which were inconsistent with the Ward identities or the decoupling of
zero-norm states (ZNS) \cite{ChanLee,ChanLee2} in the hard scattering limit to
be discussed below in Eq.(\ref{a}) to Eq.(\ref{c}).

The importance of two types of ZNS in the old covariant first quantized string
spectrum was stressed in the massive background field calculation of stringy
symmetries \cite{Lee,lee-Ov,LeePRL}. It was shown that in the weak field
approximation (but valid for all energies) an \textit{inter-particle} symmetry
transformation \cite{Lee}%
\begin{equation}
\delta C_{(\mu\nu\lambda)}=\frac{1}{2}\partial_{(\mu}\partial_{\nu}%
\theta_{\lambda)}^{2}-2\eta_{(\mu\nu}\theta_{\lambda)}^{2},\delta C_{[\mu\nu
]}=9\partial_{\lbrack\mu}\theta_{\nu]}^{2};\partial^{\mu}\theta_{\mu}%
^{2}=0,(\partial^{2}-4)\theta_{\mu}^{2}=0 \label{01}%
\end{equation}
for two propagating states $C_{(\mu\nu\lambda)}$ and $C_{[\mu\nu]}$ at mass
level $M^{2}=4$ of open bosonic string can be generated by the $D_{2}$ vector
ZNS with polarization $\theta_{\mu}^{2}$ \cite{Lee}%

\begin{equation}
|D_{2}\rangle=[(\frac{1}{2}k_{\mu}k_{\nu}\theta_{\lambda}^{2}+2\eta_{\mu\nu
}\theta_{\lambda}^{2})\alpha_{-1}^{\mu}\alpha_{-1}^{\nu}\alpha_{-1}^{\lambda
}+9k_{\mu}\theta_{\nu}^{2}\alpha_{-2}^{[\mu}\alpha_{-1}^{\nu]}-6\theta_{\mu
}^{2}\alpha_{-3}^{\mu}]\left\vert 0,k\right\rangle ,\text{ \ }k\cdot\theta
^{2}=0.\label{02}%
\end{equation}
Incidentally, a set of discrete ZNS $G_{J,M}^{+}$ \cite{ChungLee1} were
constructed and shown to form the $w_{\infty}$ spacetime symmetry algebra of
the toy $2D$ string theory%
\begin{equation}
\int\frac{dz}{2\pi i}G_{J_{1},M_{1}}^{+}(z)G_{J_{2},M_{2}}^{+}(0)=(J_{2}%
M_{1}-J_{1}M_{2})G_{J_{1}+J_{2}-1,M_{1}+M_{2}}^{+}(0).\label{2D2}%
\end{equation}

All these results strongly suggest that ZNS are crucial to study high-energy
stringy symmetry. The first set of linear relations among hard SSA was
obtained in \cite{ChanLee,ChanLee2} for the mass level $M^{2}=4$ of the $26D$
open bosonic string theory by the method of decoupling of ZNS. (Note that the
decoupling of ZNS was also used in the group theoretical calculation of SSA to
fix the measure in the SSA calculation \cite{NW5}). By solving the following
three linear relations derived from the decoupling of high-energy ZNS or
stringy Ward identities \cite{JCLee} among the four leading order hard SSA
\cite{ChanLee,ChanLee2}%

\begin{align}
\mathcal{T}_{LLT}^{5\rightarrow3}+\mathcal{T}_{(LT)}^{3}  &  =0,\label{a}\\
10\mathcal{T}_{LLT}^{5\rightarrow3}+\mathcal{T}_{TTT}^{3}+18\mathcal{T}%
_{(LT)}^{3}  &  =0,\label{b}\\
\mathcal{T}_{LLT}^{5\rightarrow3}+\mathcal{T}_{TTT}^{3}+9\mathcal{T}%
_{[LT]}^{3}  &  =0, \label{c}%
\end{align}
one obtains the ratios \cite{ChanLee,ChanLee2}%
\begin{equation}
\mathcal{T}_{TTT}:\mathcal{T}_{LLT}:\mathcal{T}_{(LT)}:\mathcal{T}%
_{[LT]}=8:1:-1:-1. \label{03}%
\end{equation}
These ratios were justified by a set of sample calculation of hard SSA
\cite{ChanLee,ChanLee2}. Similar results were obtained for the mass level
$M^{2}=6 $ \cite{ChanLee,ChanLee2}. On the other hand, A remedy calculation
was performed in \cite{CHL} to recover the missing terms calculated in
Eq.(\ref{d}) \cite{GrossManes} in order to obtain the correct four ratios in
Eq.(\ref{03}).

The results in Eq.(\ref{03}) above can be generalized to arbitrary mass levels
$M^{2}=2(N-1)$ \cite{CHLTY2,CHLTY1}
\begin{equation}
\frac{T^{(N,2m,q)}}{T^{(N,0,0)}}=\left(  -\frac{1}{M}\right)  ^{2m+q}\left(
\frac{1}{2}\right)  ^{m+q}(2m-1)!!.\label{04}%
\end{equation}
In addition to the method of decoupling of ZNS, a dual method called the
Virasoro constraint method and a corrected saddle-point calculation (for
string-tree amplitudes) also gave the same result in Eq.(\ref{04}). See the
recent review papers \cite{review,over}. It is important to note that the
linear relations and ratios obtained by the decoupling of ZNS are valid for
all string-loop orders since ZNS should be decouped for all loop amplitudes
due to unitarity of the theory. This important fact was not shared by the
saddle-point calculation in \cite{GM1,GrossManes} and neither of the
calculation performed in \cite{West2}. On the other hand, one believes that by
keeping $M$ fixed as a finite constant in the ZNS calculation one can obtain
more information about the high-energy behavior of string theory in constrast
to the tensionless string ($\alpha^{\prime}\rightarrow\infty$) approach
\cite{less12} in which all string states are massless.

Since the linear relations obtained by the decoupling of ZNS are valid order
by order and share the same forms for all orders in string perturbation
theory, one expects that there exists \textit{stringy symmetry} of the theory.
Indeed, Two such symmetry groups were suggested recently to be the $SL(5,C)$
group in the \textit{Regge} scattering limit \cite{review,over} and the
$SL(4,C)$ group in the \textit{Non-relativistic} scattering limit
\cite{review,over}. Moreover, it was shown that the ratios in Eq.(\ref{04})
can be extracted from the Regge SSA \cite{review,over}.

More recently, the authors in \cite{LSSA} constructed the exact SSA of three
tachyons and one \textit{arbitrary} string state, or the Lauricella SSA (LSSA)
\cite{LLY2}
\begin{align}
A_{st}^{(r_{n}^{T},r_{m}^{P},r_{l}^{L})} &  =\prod_{n=1}\left[  -(n-1)!k_{3}%
^{T}\right]  ^{r_{n}^{T}}\cdot\prod_{m=1}\left[  -(m-1)!k_{3}^{P}\right]
^{r_{m}^{P}}\prod_{l=1}\left[  -(l-1)!k_{3}^{L}\right]  ^{r_{l}^{L}%
}\nonumber\\
&  \cdot B\left(  -\frac{t}{2}-1,-\frac{s}{2}-1\right)  F_{D}^{(K)}\left(
-\frac{t}{2}-1;R_{n}^{T},R_{m}^{P},R_{l}^{L};\frac{u}{2}+2-N;\tilde{Z}_{n}%
^{T},\tilde{Z}_{m}^{P},\tilde{Z}_{l}^{L}\right)  \label{st1}%
\end{align}
in the $26D$ open bosonic string theory. In addition, they discovered the Lie
algebra of the $SL(K+3,C)$ symmetry group \cite{Group,slkc}%
\begin{equation}
\left[  \mathcal{E}_{ij},\mathcal{E}_{kl}\right]  =\delta_{jk}\mathcal{E}%
_{il}-\delta_{li}\mathcal{E}_{kj};\text{ \ }1\leqslant i,j\leqslant K+3
\end{equation}
valid for \textit{all} kinematic regimes of the LSSA. It was further shown
that all the LSSA can be solved \cite{solve} by the recurrence relations
associated with the $SL(K+3,C)$ group and expressed in terms of one amplitude.
Moreover, the ratios presented in Eq.(\ref{04}) can be rederived by the LSSA
in Eq.(\ref{st1}) in the hard scattering limit \cite{LSSA}.


\begin{thebibliography}{99}                                                                                               %


\bibitem {Gross}David~J. Gross. \newblock {High-Energy Symmetries of String
Theory}. \newblock {\em Phys. Rev. Lett.}, 60:1229, 1988.

\bibitem {Gross1}D.~J. Gross.
\newblock {Strings at superPlanckian energies: In search of the string
	symmetry}. \newblock In \emph{{In *London 1988, Proceedings, Physics and
mathematics of strings* 83-95. (Philos. Trans. R. Soc. London A329 (1989)
401-413).}}, 1988.

\bibitem {GM}David~J Gross and Paul~F Mende.
\newblock {The high-energy behavior of string scattering amplitudes}.
\newblock {\em Phys. Lett. B}, 197(1):129--134, 1987.

\bibitem {GM1}David~J Gross and Paul~F Mende. \newblock {String theory
beyond the Planck scale}. \newblock {\em Nucl. Phys. B}, 303(3):407--454, 1988.

\bibitem {GrossManes}David~J Gross and JL~Manes. \newblock {The high energy
behavior of open string scattering}. \newblock {\em Nucl. Phys. B},
326(1):73--107, 1989.

\bibitem {Polchin}J. Polchinski, "String Theory", Sec. 9.8, Vol. I, Cambridge
University Press, 1998.

\bibitem {Kaku}M. Kaku, "Strings,Conformal Fields, and Topology: An
Introduction", Sec.9.4, Springer-Verlag, 1991.

\bibitem {ChanLee}Chuan-Tsung Chan and Jen-Chi Lee. \newblock {Stringy
symmetries and their high-energy limits}. \newblock {\em Phys. Lett. B},
611(1):193--198, 2005.

\bibitem {ChanLee2}Chuan-Tsung Chan and Jen-Chi Lee. \newblock {Zero-norm
states and high-energy symmetries of string theory}. \newblock {\em Nucl.
Phys. B}, 690(1):3--20, 2004.

\bibitem {CHL}Chuan-Tsung Chan, Pei-Ming Ho, and Jen-Chi Lee.
\newblock {Ward identities and high energy scattering amplitudes in string
	theory}. \newblock {\em Nucl. Phys. B}, 708(1):99--114, 2005.

\bibitem {Closed}Chuan-Tsung Chan, Jen-Chi Lee, and Yi~Yang.
\newblock {Notes on high-energy limit of bosonic closed string scattering
	amplitudes}. \newblock {\em Nucl. Phys. B}, 749(1):280--290, 2006.

\bibitem {bcj0}The first BCJ relation was the discovery of the \textit{hard}
string BCJ derived in \cite{Closed} (2006), which was earlier than the field
theory BCJ (2008) and string BCJ (2009).

\bibitem {KLT}H.~Kawai, David~C Lewellen, and S-HH Tye.
\newblock {A relation between tree amplitudes of closed and open strings}.
\newblock {\em Nucl. Phys. B}, 269(1):1--23, 1986.

\bibitem {West2}Nicolas Moeller and Peter West. \newblock {Arbitrary four
string scattering at high energy and fixed angle}. \newblock {\em Nucl.
Phys. B}, 729(1):1--48, 2005.

\bibitem {NW5}P. C. West, A Brief Review of the Group Theoretic Approach to
String Theory, in \textquotedblright Conformal Field Theories and Related
Topics\textquotedblright, Nucl. Phys. B (Proc. Suppl) 5B (1988) 217.

\bibitem {Lee}Jen-Chi Lee. \newblock New symmetries of higher spin states in
string theory. \newblock {\em Physics Letters B}, 241(3):336--342, 1990.

\bibitem {lee-Ov}Jen-Chi Lee and Burt~A Ovrut.
\newblock {Zero-norm states and enlarged gauge symmetries of the closed bosonic
	string with massive background fields}. \newblock {\em Nucl. Phys. B},
336(2):222--244, 1990.

\bibitem {LeePRL}Jen-Chi Lee.
\newblock {Decoupling of degenerate positive-norm states in string theory}.
\newblock {\em Phys. Rev. Lett.}, 64(14):1636, 1990.

\bibitem {ChungLee1}Tze-Dan Chung and Jen-Chi Lee.
\newblock {Discrete gauge states and $w_{\infty}$ charges in c= 1 2D gravity}.
\newblock {\em Phys. Lett. B}, 350(1):22--27, 1995.

\bibitem {JCLee}Jen-Chi Lee.
\newblock {Generalized on-shell Ward identities in string theory}.
\newblock {\em Prog. of th. phys.}, 91(2):353--360, 1994.

\bibitem {CHLTY2}Chuan-Tsung Chan, Pei-Ming Ho, Jen-Chi Lee, Shunsuke
Teraguchi, and Yi~Yang. \newblock {High-energy zero-norm states and
symmetries of string theory}. \newblock {\em Phys. Rev. Lett.}, 96(17):171601, 2006.

\bibitem {CHLTY1}Chuan-Tsung Chan, Pei-Ming Ho, Jen-Chi Lee, Shunsuke
Teraguchi, and Yi~Yang.
\newblock {Solving all 4-point correlation functions for bosonic open string
	theory in the high-energy limit}. \newblock {\em Nucl. Phys. B},
725(1):352--382, 2005.

\bibitem {review}Jen-Chi Lee and Yi~Yang. \newblock Review on high energy
string scattering amplitudes and symmetries of string theory. \newblock {\em
arXiv preprint arXiv:1510.03297}, 2015.

\bibitem {over}Jen-Chi Lee and Yi~Yang. \newblock Overview of high energy
string scattering amplitudes and symmetries of string theory. \newblock {\em
Symmetry}, 11(8):1045, 2019.

\bibitem {less12}Sagnotti, A., and M. Tsulaia. "On higher spins and the
tensionless limit of string theory." Nuclear Physics B 682.1-2 (2004): 83-116.

\bibitem {LSSA}S. H. Lai, J. C. Lee and Yi Yang, "Recent developments of the
Lauricella string scattering amplitudes and their exact $SL(K+3,C)$ Symmetry",
Symmetry 13 (2021) 454, arXiv:2012.14726 [hep-th].

\bibitem {LLY2}Sheng-Hong Lai, Jen-Chi Lee, and Yi~Yang. \newblock The
lauricella functions and exact string scattering amplitudes.
\newblock {\em Journal of High Energy Physics}, 2016(11):62, 2016.

\bibitem {Group}S. H. Lai, J. C. Lee and Yi Yang, "The $SL(K+3,C)$ Symmetry of
the Bosonic String Scattering Amplitudes", Nucl. Phys. B 941 (2019) 53-71.

\bibitem {slkc}Willard Miller~Jr. \newblock Symmetry and separation of
variables. Addison-Wesley, Reading, Massachusetts, \newblock1977.

\bibitem {solve}S. H. Lai, J. C. Lee and Yi Yang, "Solving Lauricella String
Scattering Amplitudes through recurrence relations", JHEP 09 (2017) 130.
\end{thebibliography}
\end{document}